\documentclass[aps,prl,twocolumn,superscriptaddress,showpacs]{revtex4}

\usepackage{epsf,graphicx,amssymb,subfigure,amsmath}

\newcommand{\be}{\begin{equation}}  
\newcommand{\ee}{\end{equation}}  
\newcommand{\ba}{\begin{eqnarray}}  
\newcommand{\ea}{\end{eqnarray}}  
\newcommand{\bi}{\begin{itemize}}  
\newcommand{\ei}{\end{itemize}}  
\newcommand{\bc}{\begin{center}}  
\newcommand{\ec}{\end{center}}  

\begin{document}

\title{Sublattice synchronization of chaotic networks with delayed couplings}

\author{Johannes Kestler}
\affiliation{Institute for Theoretical Physics, University of W\"urzburg, Am Hubland, 97074 W\"urzburg, Germany}
\author{Wolfgang Kinzel}
\affiliation{Institute for Theoretical Physics, University of W\"urzburg, Am Hubland, 97074 W\"urzburg, Germany}
\author{Ido Kanter}
\affiliation{Department of Physics, Bar-Ilan University, Ramat-Gan, 52900 Israel}

\date{March 2, 2007}

\begin{abstract}

  Synchronization of chaotic units coupled by their time delayed
  variables are investigated analytically.  A new type of cooperative
  behavior is found: sublattice synchronization. Although the units of
  one sublattice are not directly coupled to each other, they
  completely synchronize without time delay.  The chaotic trajectories
  of different sublattices are only weakly correlated but not related
  by generalized synchronization. Nevertheless, the trajectory of one
  sublattice is predictable from the complete trajectory of the other
  one. The spectra of Lyapunov exponents are calculated analytically
  in the limit of infinite delay times, and phase diagrams are derived
  for different topologies.
\end{abstract}

\pacs{05.45.Xt, 05.45.Ra, 05.45.--a}

\maketitle

Two identical chaotic systems which are coupled to each other can
synchronize to a common chaotic trajectory. Both of the systems are
chaotic, but after synchronization the two chaotic trajectories are
locked to each other \cite{Pikovsky:book, Schuster:2005}. This
phenomenon has attracted a lot of research, partly because chaos
synchronization has the potential to be applied for novel secure
communication systems \cite{Pecora:1990, Cuomo:1993}.  In fact,
synchronization and bit exchange with chaotic semiconductor lasers has
recently been demonstrated over a distance of 120 km in a public
fiber-optic network \cite{Argyris:2005}.

Typically, the coupling between chaotic units has a time delay due to
transmission of the exchanged signal. Nevertheless, two chaotic
systems can synchronize without delay, isochronically, although the
delay time may be very long. This counterintuitive phenomenon has
recently been demonstrated with chaotic semiconductor lasers
\cite{Klein:06:PRE73, Fischer:2006:PRL}, and it is discussed in the
context of corresponding observations on correlated neuronal activity
\cite{Adrian:2006:SCI, Engel:1991:SCI, Campbell:1998}.

\begin{figure}[ht]
        \includegraphics[width=.4\textwidth]{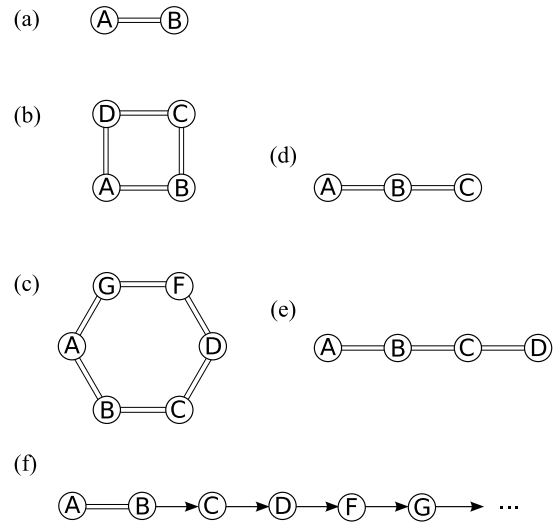}
        \caption{Small networks of chaotic units. Double lines signify
          bidirectional couplings whereas arrows show unidirectional
          couplings. It turns out that each ring with an even number
          $N$ of units has Lyapunov exponents which are identical to a chain with $N/2+1$  units.} 
        \label{fig:anordnungen}
\end{figure}

In this Letter we give an analytic solution of small networks of
chaotic units coupled by their time-delayed variables.  Motivated by
chaotic lasers, we consider the competition between delayed
self-feedback and coupling.  For regular networks which can be
decomposed into two identical sublattices -- e.g. a ring with $N=4$
units in Fig.~\ref{fig:anordnungen}(b) -- a new kind of
synchronization is observed: The units (A, C) as well as the units (B,
D) are completely synchronized. Even for arbitrarily long delay times,
the synchronization is isochronical, i.e. the trajectories of units A
and C are identical without any time-shift.  Although synchronization
of (A, C) is caused by the trajectory of (B, D) and vice versa, the
trajectories of A and B are not synchronized.  Actually, we do not
even find generalized synchronization between A and B.  The
cross-correlations are symmetric, there is neither a leader nor a
laggard for the two sublattices.  Each unit is passive, i.e. it has
negative Lyapunov exponents when it is isolated and driven by an
external signal. Nevertheless, the mutual interaction of passive
elements leads to high-dimensional chaos with synchronized units in
each sublattice.

We have found  sublattice synchronization in several chaotic networks,
in particular for  rate equations describing chaotic semiconductor lasers
\cite{Fischer:2222}. In this Letter we analyze sublattice synchronization using a simpler
system, namely chaotic Bernoulli maps with delay, which allows analytical as
well as comprehensive numerical investigations. In the limit of long
delay times, the spectra of Lyapunov exponents, which are calculated
analytically, determine the phase boundaries between sublattice synchronization, complete
synchronization  and complete chaos.

Two mechanisms are included in our model: 1. Delayed feedback and 2.
delayed mutual coupling. The first mechanism can generate high
dimensional chaos for each single unit whereas the second one
synchronizes the network.  The phenomena caused by these two
mechanisms are captured by a network of iterated maps, given by the
following equations, $j, k = 1, \ldots, N$:
\begin{multline}
        x_t^j = (1-\varepsilon) f(x_{t-1}^j) + {}\\ 
        {} + \varepsilon \Big[ \kappa f(x_{t-\tau}^j) 
                + (1-\kappa) \sum_{k} F_{j,k} \; f(x_{t-\tau}^k) \Big]  \label{eins} 
\end{multline}

Each node $j$ of the network contains a variable $x^j_t \in [0,1]$
which is iterated in time $t$.  $F_{j,k}$ is the weighted adjacency
matrix which has a component $1/z_j$ if node $j$ is driven by node $k$
and 0 otherwise, and $z_j$ is the number of connections to node $j$.

Analytic calculations are
possible for the Bernoulli shift map $f(x) = (ax)\; \mathrm{mod} \,1$
which is chaotic for $a > 1$ \cite{Lepri:1993:PHD}. $\tau$ is the
delay time, which, for simplicity, is identical for the feedback as well
as for the coupling. The parameter
$\varepsilon$ determines the strength of the delay terms while
$\kappa$ determines the relative strength of the self-feedback
compared to the mutual coupling terms.

Now let us discuss stationary  solutions of (\ref{eins}).
Complete synchronization, $x^j_t = x^k_t$ for all $(j, k)$, is a
solution of (\ref{eins}). 
In addition, several kinds of incomplete synchronization are  solutions, as
well. For example, sublattice synchronization,   
$x^j_t = x^k_t$ for all nodes $(j, k)$ in each sublattice, is a solution,
too, if the network can be decomposed into two interconnecting
sublattices, as for the chains and rings in Fig.~\ref{fig:anordnungen}. But are these
solutions  stable? Their stability
 has to be determined from equation (\ref{eins}). 
Let $\delta x^j_t$ be a small deviation from any given trajectory $x^j_t$. 
The linearized equations (\ref{eins}) are solved with the ansatz $\delta
x^j_t = c^t \; v_j$, where the value $\lambda = \ln |c|$ is
the corresponding Lyapunov exponents. We find that 
the vectors $(v_j)_{j=1,..,N}$ are eigenvectors of the adjacency
matrix $F$, and their corresponding eigenvalues $\mu_q, q=0,...,N-1$
determine the Lyapunov exponents by

\be
\mu_q = -\frac{(1 - \varepsilon) a c^{\tau - 1} + \varepsilon \kappa a - c^{\tau} }{ \varepsilon (1 - \kappa) a }
\label{mu}
\ee

\begin{figure}[ht]
  \includegraphics[width=.3\textwidth]{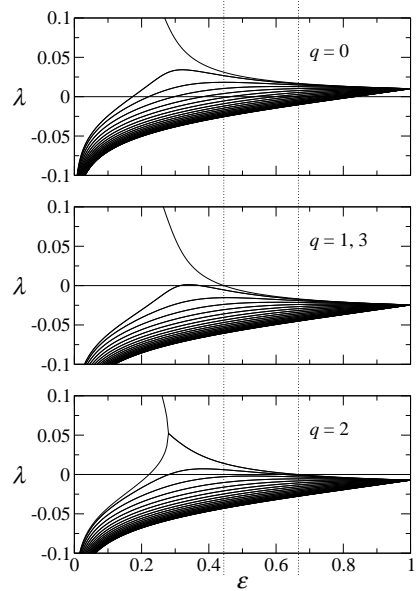}
  \caption{Lyapunov spectra for the different modes $q$ for $N = 4, a =
    1.5, \tau = 40, \kappa = 1/4$. At $\varepsilon = 4/9$ the ($q =
    1$) mode becomes stable ($\to$ sublattice synchronization). At
    $\varepsilon = 2/3$ the ($q = 2$) mode, $(+1,-1,+1,-1)$, becomes
    stable, too ($\to$ complete synchronization). See also
    Fig.~\ref{fig:gebiete}}
        \label{fig:lyapunov}
\end{figure}

For each eigenvector, indexed by $q=0,...,N-1$, 
this polynomial equation (\ref{mu}) for $c$ has $\tau$ complex solutions $c_{q,m}\, ,
m = 1, \ldots, \tau$ giving a total of $ N \tau$ Lyapunov exponents
which are plotted in Fig. \ref{fig:lyapunov}. 
The eigenvectors of $F$ determine the $N$ directions of perturbation
$v_{q,j}$ which shrink or explode with the corresponding values
of $c$.

Here we are interested in long delay times, $\tau \to \infty$. It
turns out that for each eigenvalue $\mu_q$, $q=0, \ldots ,N-1$, there
is one Lyapunov exponent of order one, and $\tau - 1$ exponents of
order $1/\tau$ (see Fig. \ref{fig:lyapunov}). The large exponent is given by $\lambda ^1_q = \ln
[(1 - \varepsilon)a]$ and merges with the band of the other $\tau - 1$
exponents at the critical point 
\be 
\varepsilon_c = \frac{a-1}{a} .
\ee 
Consequently, for $\varepsilon < \varepsilon_c$, stable synchronization does not
exist.

\begin{figure}[ht]
  \includegraphics[width=.3\textwidth]{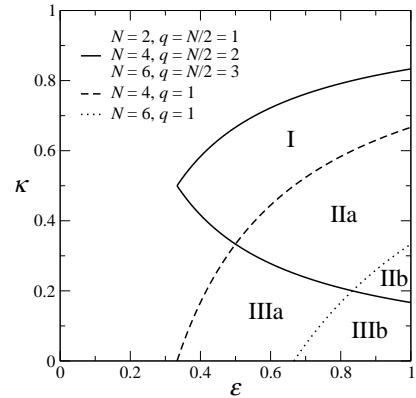}
  \caption{Phase diagram for $a = 1.5$. See text for assignment of the
    different regions to complete and sublattice synchronization.}
        \label{fig:gebiete}
\end{figure}

The eigenvalues of the adjacency matrix $F$ determine the stability of
synchronized trajectories according to (\ref{mu}), and they depend on
the topology of the lattice.  Let us consider the rings and chains
depicted in Fig.~\ref{fig:anordnungen}. We find that the phase diagram
of all six topologies is given by Fig.~\ref{fig:gebiete}. To see this,
let us consider a ring of $N$ mutually coupled chaotic units with
self-feedback, described by (\ref{eins}).
Figure~\ref{fig:anordnungen} shows $N =$ 2, 4 and 6. By symmetry, the
eigenvectors $v_{q,j}$ and eigenvalues $\mu_q$ of the adjacency matrix
$F$ are the Fourier modes, ($q=0,...,N-1$) 
\be v_{q,j} = \exp ( 2 \pi
{\mathrm i} \, j \, q/N),\quad \mu_q = \cos \left( \frac{2 \pi q}{N}
\right) .  
\ee 
These eigenvalues determine the Lyapunov spectrum by
(\ref{mu}).  Using the mehtods of Ref.~\cite{Lepri:1993:PHD}, for
large values of $\tau$ the maximal Lyapunov exponent is calculated as
\be
\label{lambdaformel}
\lambda_q^{\mathrm{max}}=\frac{1}{\tau} \ln \left| \frac{a \varepsilon
    \left[ (1-\kappa) \cos \left( \frac{2 \pi q}{N} \right) + \kappa
    \right] }{1-a(1-\varepsilon)} \right| \quad (\varepsilon >
\varepsilon_c).  
\ee

The mode $(q=0)$ with the corresponding eigenvector $(1, 1, 1,
\ldots)$ is always unstable, $\lambda_{q=0}^{\mathrm{max}}>0$, hence
the system is chaotic for all parameters $\varepsilon$ and $\kappa$.
In fact, the system is in a state of high-dimensional chaos
(hyperchaos), since the Kaplan-Yorke dimension increases linearly with
the delay time $\tau$.

The superpositions of the  modes ($q=1,q=N-1$), which have degenerate
eigenvalues,  are perturbations 
which destroy any structure of
synchronization. They are  unstable for
\be
\kappa > \kappa_{q=1}= \frac{1-a}{a \varepsilon \left[ 1- \cos \left( \frac{2 \pi}{N} \right) \right] }+1 \; .
\label{acht}
\ee

An analysis of  (\ref{lambdaformel}) shows that for even values of $N$, the only perturbation which is
relevant in the region $\kappa < \kappa_{q = 1}$ is the perturbation
of the mode $(q = N/2)$. The network is unstable
against this mode for
\be
\kappa < \kappa_{q=N/2} = \frac{a-1}{2 a \varepsilon} \; .
\label{neun}
\ee

Since this mode corresponds to the eigenvector $(+1, -1, +1, -1, \ldots, -1)$,
it can destroy only  complete  but  not sublattice synchronization.

Equations (\ref{acht}) and (\ref{neun}) immediately determine the
phase diagram of the rings with $N$ units in the ($\kappa,\epsilon$)
plane,   Fig.~\ref{fig:gebiete}.
The ring with $N = 4$ units, Fig.~\ref{fig:anordnungen}(b), is
completely synchronized in region II (= IIa plus IIb).  But complete synchronization is
cannot exist in region III, because there  the $q = N/2$ mode is unstable.
In this region  the system is synchronized in a sublattice
configuration: the trajectory of A is completely identical, without
delay, with the one of C, and correspondingly, B is identical to D.

When increasing the number $N$ of units the $(q = 1)$ line of
Fig.~\ref{fig:gebiete} shrinks to the lower right corner. For $N = 6$
we find complete synchronization in the region IIb, while sublattice
synchronization occurs in region IIIb. In the latter region the
trajectories of (A, C, F) as well as (B, D, G) are identical, but the two
different trajectories do not synchronize.

Finally, for large system size $N$ synchronization disappears completely.
Complete synchronization exists up to
\be
        N < \frac{2\pi}{\arccos \left( \frac{3-a}{1+a} \right)} ,
\ee
while sublattice synchronization remains for larger sizes up to 
\be
        N < \frac{2\pi}{\arccos \left( \frac{1}{a} \right)} .
\ee
Consequently, even for arbitrary large rings one obtains regions of
complete and sublattice synchronization, if the chaos is sufficiently
weak. Asymptotically, sublattice synchronization is stable if $a$ is
smaller than $a \sim 1+2 (\pi/N)^2$.

In our model the rings with an even number $N$ of chaotic units are
equivalent to chains with $N/2 + 1$ units, i.e. the corresponding
rings have identical, but degenerate, Lyapunov exponents. Hence the
phase diagram of Fig.~\ref{fig:gebiete} describes synchronization of
the chains Fig.~\ref{fig:anordnungen}(a), (d) and (e), as well.
Regions II and III also describe the infinite chain with directed
bonds of Fig.~\ref{fig:anordnungen}(f). If the left unit A is just a
mirror , corresponding to a self-feedback of B with $\kappa=1$, the
chain is in state of complete synchronization in these regions. If, however, A is a chaotic
unit, the infinite chain switches to a state of complete synchronization in region II and
sublattice synchronization in region III.

These findings are not
specific to rings and chains, since we found sublattice synchronization in square lattices
with periodic boundaries (with even side lengths) as well as for free
boundaries (with even or odd side length).  In addition, we found sublattice synchronization
for a small triangular lattices with three sublattices.  Each
sublattice synchronizes completely, without time shifts, whereas the
three trajectories have only weak correlations.

Synchronization of the units of one sublattice is relayed by the
chaotic trajectory of the other one. Hence one would expect some
relation between these two chaotic trajectories (or three for the
triangular lattice). 

\begin{figure}[ht]
        \includegraphics[width=.3\textwidth]{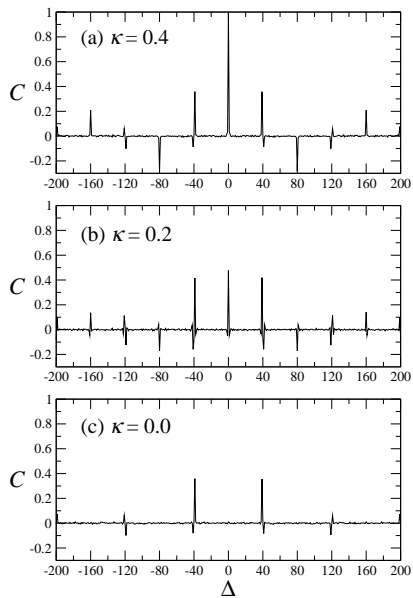}
        \caption{Cross-correlation $C$ between $A_t$ and
          $B_{t+\Delta}$ for $a = 1.5, \tau = 40, \varepsilon = 0.7$.
          (a): Complete Synchronization, (b,c): sublattice synchronization.}
        \label{fig:correlations}
\end{figure}

The cross correlations between the two sublattices are shown in Fig.
\ref{fig:correlations}.  For sublattice synchronization the central
peak, a reminder of complete synchronization, still exists, in
addition to correlations shifted by multiples of the delay time
$\tau$. Only when self-feedback is switched off, $\kappa=0$, the
isochronal correlations disappear.  The correlations are completely
symmetric in time shift $\pm \Delta$, there is no symmetry breaking to
a leader/laggard scenario.  In addition, when crossing the phase
boundary between sublattice synchronization, complete synchronization
and complete chaos, the central peak of the cross-correlation changes
discontinuously, similar to a subcritical bifurcation. For large, but
finite observation times, our simulations even show hysteresis
effects.
 
\begin{figure}[h]
  \includegraphics[width=.3\textwidth]{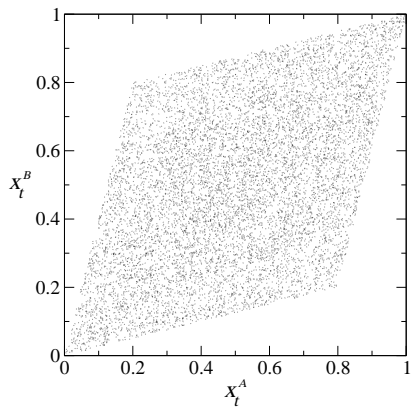}
        \caption{Two trajectories of the sublattice synchronization phase,
          for the networks of Fig.~\ref{fig:anordnungen} with $\tau =
          1, \varepsilon = 0.8, \kappa = 0, a = 3.0$. Obviously, there
          is no functional relation between the two trajectories.}
        \label{fig:GS}
\end{figure}

Apart from these weak correlations, is there any functional relation
between the two different trajectories? In particular, are the
trajectories in a state of \emph {generalized synchronization}
\cite{Abarbanel:1996:PRE}? For our model, generalized synchronization
means that there exists a function between the sequence $(x^A_t,
\ldots, x^A_ {t - \tau})$ of A and $(x^B_t, \ldots, x^B_ {t - \tau})$
of B.  This function may be fractal \cite{Pikovsky:book} or may have
several branches, which usually makes an analysis difficult.  But at
least for $\tau=1$ of $\tau=2$ the relation between these two vectors
can be easily visualized. Fig.~\ref{fig:GS} shows that the
trajectories are not related by generalized synchronization in the
phase of sublattice synchronization.

However, there is a related question: Can the trajectory of sublattice
B be predicted from the knowledge of the complete trajectory of A
\cite{Afraimovich:2002:PRE}?  This question has a positive answer for
the configurations of Fig.~\ref{fig:anordnungen}. In regions II and
III all units are passive. This means, if we omit the drive, the last
term of (\ref{eins}), each unit has negative Lyapunov exponents. Hence
if this unit is driven, it will relax to a unique trajectory. Since we
find sublattice synchronization only for passive units, we can predict the trajectory of B
from the knowledge of the \emph{complete} trajectory of A after some
transient time.  Note that for two units we find complete synchronization for active units,
as well, but in this case, region I, prediction is trivial.

Sublattice synchronization has not yet been found in experiments.
However, a chain of three lasers Fig.~\ref{fig:anordnungen}(d), has
been investigated with semiconductor lasers without feedback, $\kappa
= 0$ \cite{Fischer:2006:PRL}.  Synchronization of the outer
lasers has been observed, being relayed by a different chaotic
trajectory of the internal laser, in agreement with our model. But
there are differences to our results: In our model, the cross
correlation is symmetric in time shift, there is neither a  spontaneous
symmetry breaking nor a leader/laggard scenario\cite{Heil:2001}, and there is no generalized
synchronization between the synchronized and the connecting units \cite{Landsman:2007:PRE}.

\begin{acknowledgments}
We would like to thank J\"{o}rg Reichardt, Andreas Ruttor and Evi Kopelowitz  for useful
discussions.
\end{acknowledgments}

\bibliography{sublattice}

\end{document}